\documentclass[conference]{IEEEtran}

\usepackage[utf8]{inputenc}
\usepackage{cite}
\usepackage{amsmath,amssymb,amsfonts}
\usepackage{graphicx}
\usepackage{textcomp}
\usepackage{xcolor}
\usepackage{booktabs}
\usepackage{url}      
\usepackage{algorithm}
\usepackage{algpseudocode}

\usepackage{algorithm}

\begin{document}

\title{Hybrid MAC Protocol with Integrated Multi-Layered Security for Resource-Constrained UAV Swarm Communications
}

\author{\IEEEauthorblockN{Dhrumil Bhatt\textsuperscript{*}}
\IEEEauthorblockA{Department of Electronics and\\ Electrical Engineering\\
Manipal Institute of Technology\\
Manipal Academy of Higher Education\\
Manipal, India\\
dhrumil.bhatt@gmail.com}
\and
\IEEEauthorblockN{Siddharth Penumatsa\textsuperscript{*}}
\IEEEauthorblockA{Department of Electronics and \\Communication Engineering\\
Manipal Institute of Technology\\
Manipal Academy of Higher Education\\
Manipal, India\\
psiddharthv06@gmail.com}
\and
\IEEEauthorblockN{Vidushi Kumar\textsuperscript{*}}
\IEEEauthorblockA{Department of Electronics and \\Electrical Engineering\\
Manipal Institute of Technology\\
Manipal Academy of Higher Education\\
Manipal, India\\
vidushi.kumar705@gmail.com}

\thanks{\textsuperscript{*}Authors contributed equally to this work.}
} 
 \IEEEoverridecommandlockouts

\maketitle

\begin{abstract}
Flying Ad Hoc Networks (FANETs) present unique challenges due to high node mobility, dynamic topologies, and stringent resource constraints. Existing routing protocols often optimize for a single metric, such as path length or energy, while neglecting the complex dependencies between network performance, security, and MAC-layer efficiency. This paper introduces a novel hardware-software co-design framework for secure and adaptive UAV swarm communications, featuring an energy-aware protocol stack. The architecture employs a multi-cast, clustered organization where routing decisions integrate dynamic trust scores, historical link quality, and internodal distance. A hybrid MAC protocol combines contention-based and scheduled channel access for optimized throughput. Security is ensured through a zero-trust model fusing cryptographic authentication with a behavioral reputation system, alongside hardware-accelerated AES-GCM encryption. Comparative analysis in an NS-3 simulation environment demonstrates the framework’s superiority in packet delivery ratio, latency, resilience, and overhead, designed to provide a scalable foundation for high-performance swarm operations.
\end{abstract}


\section{Introduction}

The rise of Flying Ad-Hoc Networks (FANETs) has redefined aerial communication paradigms, enabling autonomous, multi-UAV systems to perform collaborative missions in military, disaster response, and surveillance operations. These networks, built from rapidly mobile, self-organizing aerial nodes, face unique challenges including volatile topologies, constrained energy budgets, and severe susceptibility to cyber-physical threats \cite{diockou2024review,chen2025multidomain}.

Security remains a cornerstone issue in FANETs due to the open wireless medium, the broadcast nature of drone communications, and the lack of centralized infrastructure. While ground-based ad hoc networks have mature solutions, FANETs introduce complexities in mobility models, energy-aware operations, and cross-layer protocol interactions that render traditional approaches insufficient \cite{piekarski2025uncrewed}.

Despite advancements in routing, clustering, and trust management, most existing approaches optimize isolated layers without considering the feedback mechanisms across the MAC, network, and application layers. Recent frameworks like those integrating blockchain for secure cluster head (CH) election or fuzzy logic-based trust evaluation have shown promise but often impose heavy computational loads unsuitable for embedded platforms in UAVs \cite{zhao2021abcblockchain,prabha2018fuzzytrust}. Furthermore, studies like TCSFANET \cite{kundu2022tcsfanet} and early secure clustering protocols for WSNs \cite{han2011secureclustering} neglect critical cross-layer adaptability, cryptographic trust validation, and MAC efficiency under adversarial stress.

This paper introduces a holistic, lightweight communication architecture for UAV swarms operating under resource constraints. The proposed system combines a Dynamic Weighted Clustering Protocol, Hybrid MAC design (SC-HybridMAC), and a dual-layer security model blending cryptographic assurance with behavioral reputation scoring. Designed for real-time, scalable operations, this architecture is rigorously validated in an NS-3 simulation environment using realistic physical-layer and mobility models, thus addressing both the theoretical and implementation gaps in secure FANET design.
\section{Related Work}

Flying Ad-Hoc Networks (FANETs) represent a specialized and rapidly evolving subset of mobile ad hoc networks, where aerial nodes such as unmanned aerial vehicles (UAVs) form dynamic, self-organizing topologies. These networks face unique challenges, including high node mobility, constrained energy resources, variable link quality, and heightened susceptibility to security breaches. While numerous studies have investigated isolated components such as secure routing, clustering, or trust management, integrated and simulation-validated frameworks remain limited. 

Zhao et al. \cite{zhao2021abcblockchain} introduced a blockchain-enhanced secure routing architecture tailored for FANETs. Their protocol utilizes an improved artificial bee colony (IABC) algorithm to optimize cluster head (CH) selection based on a multi-criteria fitness function, incorporating residual energy, mobility, connectivity, online time, and trust. The scheme integrates a novel consensus mechanism, AI-PoWCA, to prevent 51\% attacks while maintaining lightweight blockchain operations. Simulation results demonstrate improvements in packet delivery ratio (PDR), throughput, and resilience. However, the reliance on blockchain infrastructure and cryptographic mining introduces significant complexity and computational cost, potentially limiting scalability in resource-constrained UAV platforms.

Prabha and Jeyanthi \cite{prabha2018fuzzytrust} proposed a fuzzy logic-based secure routing algorithm that fuses trust evaluation with energy-aware clustering. Their model evaluates trust through a three-tier system—historical, behavioral, and neighbor trust—and combines it with a fuzzy clustering mechanism for CH election. The system incorporates CH rotation and fuzzy rule inference to mitigate the effects of topology volatility. While the approach improves security and reduces re-affiliation overhead, it remains focused on routing-layer optimizations without considering adaptive medium access or physical layer variability, which are critical in dense UAV swarms.

The TCSFANET scheme by Kundu et al. \cite{kundu2022tcsfanet} advances trust-based communication by establishing cluster heads based on computed trust scores. The architecture emphasizes inter-UAV coordination and base station reliability through dynamic trust filtering of malicious UAVs. It employs a centralized cluster management approach where secure cluster leaders manage gateway transmissions. Despite its strengths in malicious node detection, TCSFANET lacks cross-layer optimization and does not model the impact of mobility or energy constraints under large-scale simulations.

Han et al. \cite{han2011secureclustering} proposed a patented secure routing and clustering method for wireless sensor networks (WSNs), emphasizing trust-based CH selection and intrusion detection with energy prediction. The approach ensures reliable communication by isolating hostile nodes and maintaining a secure cluster structure. However, its applicability is confined to static or low-mobility WSNs, and the model lacks extensions to support high-mobility, rapidly shifting topologies inherent in FANETs.

Lalouani proposes Sec-PUF, which focuses on a hardware-centric approach that leverages Physical Unclonable Functions (PUFs) as the primary security mechanism \cite{10187758}. The system architecture consists of UAVs equipped with PUFs and a trusted ground station that serves only for initialisation, without requiring persistent connectivity during operation. Sec-PUF does not propose a specific MAC protocol but rather focuses on the security layer that can be implemented over existing communication standards . Employing a four-phase security approach, authenticating Hardware fingerprint using PUF challenge-response pairs first, then Dynamic shuffling based on chaotic maps for message encryptions. Chaotic map-generated authentication tokens are generated, and the Chinese Remainder Theorem (CRT) is used for key management. The Hybrid MAC system uses a Beta probability distribution model to track node behaviour, and includes sophisticated time-decay mechanisms to ensure trust scores reflect current behaviour rather than historical performance. Sec-PUF implements hardware-based trust through PUF responses, where authentication relies on the unique, unclonable hardware characteristics of each UAV. Trust is inherently tied to the physical device rather than behavioural patterns. Sec-PUF does not implement explicit clustering mechanisms but rather focuses on swarm-wide group communication where all legitimate UAVs can establish secure communication links with any other legitimate member. Sec-PUF introduces a group key generation mechanism using the Chinese Remainder Theorem. The ground station combines UAV PUF responses to generate a group key that each individual UAV can recover using only its own hardware fingerprint.

Zhen Wang et al. adopts a simpler cluster-based model where any UAV can serve as the central control node, issuing commands to other cluster members \cite{9463621}. The Ground Control Station (GCS) is only trusted during initialization for key and parameter distribution. The architecture emphasizes minimal protocol changes for compatibility while maintaining dynamic topology management and relies on minimal extensions to MAVLink 2 without cross‐layer feedback. The system lacks any mechanism to detect misbehavior once cryptographic authentication succeeds; SCHybrid-Mac remedies this with a Beta Reputation System that continuously scores node behavior. The system proposed by  Wangextends MAVLink 2 messages but does not improve channel access efficiency and its single‐layer security uses ECIES with FourQ and precomputation for efficiency but cannot handle behavioral attacks.

These studies demonstrate meaningful advances in trust modeling, secure routing, and clustering protocols. Yet, they generally exhibit one or more of the following limitations: (i) layer-specific optimization without cross-layer feedback; (ii) lack of real-time adaptability in MAC scheduling or contention handling; (iii) evaluation in small-scale or static topologies; and (iv) omission of integrated cryptographic and trust-based decision mechanisms across layers. Furthermore, few studies adopt simulation environments that incorporate realistic mobility, fading models, energy dissipation, and adversarial threats simultaneously.

In light of these gaps, this paper presents a cross-layer framework that unifies trust-aware clustering, cryptographic security, and hybrid MAC scheduling, evaluated in a scalable NS-3 environment with realistic physical-layer and energy models. By integrating mechanisms across protocol layers and validating under adversarial conditions, the proposed system offers a more comprehensive and deployable approach to secure and resilient UAV swarm communication.

\section{Dynamic Weighted Clustering Protocol}

The Dynamic Weighted Clustering Protocol is engineered to create an efficient, resilient, and scalable network structure for a swarm of UAVs. It moves beyond simplistic selection metrics by employing a multi-criteria algorithm to elect the most suitable Cluster Heads (CHs). This ensures that the nodes chosen to manage clusters are not just conveniently located, but are also the most reliable and robust for the role.

\begin{figure}[ht]
\centering
\includegraphics[width=0.4\textwidth]{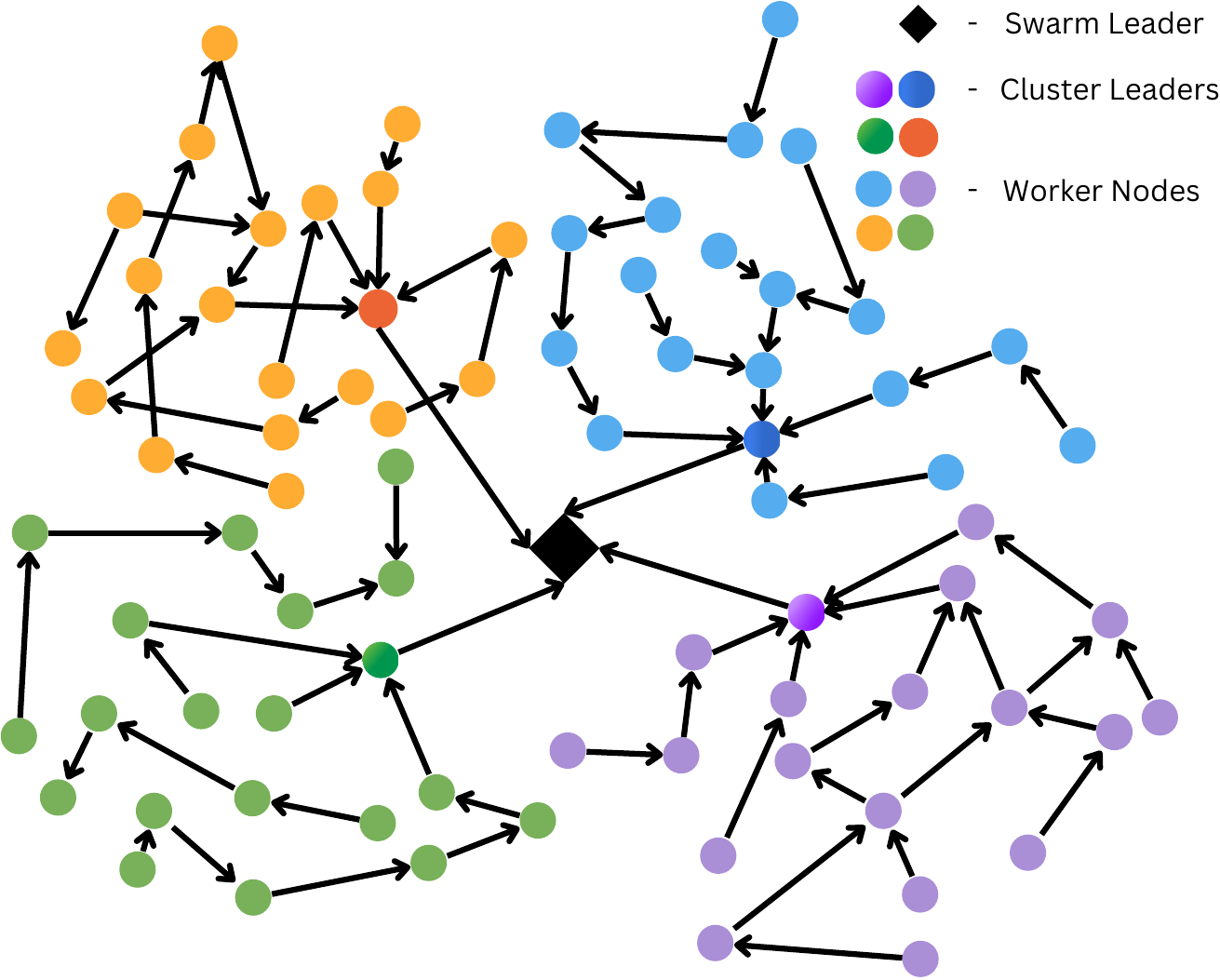}
\caption{The figure above represents a clustered UAV swarm topology generated by the proposed framework, where each colored group represents a cluster with a designated Cluster Leader. The Swarm Leader centrally coordinates inter-cluster communication using SC-HybridMAC and trust-aware routing.}
\label{fig:swarm}
\end{figure}

\subsection{Cluster Head Election Algorithm}

The core of the CH election process is the calculation of a candidacy score, $S_i$, for each node $i$ in the network. A node with a higher score is a more qualified candidate for the CH role. The election is won by the node with the highest score within a given locality.

The formula for the candidacy score is:
\begin{equation}
    S_i = w_T \cdot T_i + w_E \cdot E_i + w_C \cdot C_i
    \label{eq:ch_score_expanded}
\end{equation}

This equation represents a weighted sum, where each component contributes to the justification of the node to be a CH. 

The strength of this algorithm lies in its adaptability, controlled by the weights $w_T$ (trust), $w_E$ (energy), and $w_C$ (connectivity). These weights are not fixed; they are configurable parameters set at the beginning of a mission to align the network's behavior with the mission's primary objectives. The sum of the weights typically equals 1 ($w_T + w_E + w_C = 1$).

\subsubsection{Normalized Trust Score ($T_i$)}
 The trust score is a dynamically calculated metric that represents the reliability and behavioral integrity of a node. It is derived from the Beta Reputation System, where past interactions are used to predict future behavior. 
In a potentially adversarial environment, a node's willingness and ability to follow protocol rules are paramount. A high trust score indicates that the node has a history of positive interactions. 

\subsubsection{Normalized Residual Energy ($E_i$)}
 This parameter represents the amount of battery life remaining for node $i$. To be used in the weighted sum, this absolute value is normalized, by dividing the node's current energy by its maximum initial energy. This yields a value between 0 (empty) and 1 (full).
The role of a CH is more energy-intensive than that of a regular member node. If a node with low energy is elected as a CH, it will quickly deplete its power. This triggers a new election, causing network-wide overhead and a period of instability. Prioritizing nodes with high residual energy leads to longer-lasting clusters and greater overall network lifetime.

\subsubsection{ Degree of Connectivity ($C_i$)}
 The degree of connectivity, also known as node degree, is a measure of how many other nodes a given node $i$ can communicate with directly. It is a count of the node's one-hop neighbors. This value is also normalized by dividing by the total number of nodes in the swarm minus one ($N-1$), to scale it to the [0, 1] range.
A higher degree of connectivity means the CH can directly reach more nodes within its cluster, reducing the number of hops required for intra-cluster communication. This leads to lower latency and higher reliability. 

\section{Protocol Operations and Security}

\subsection{Cluster Maintenance}
The network topology in a UAV swarm is unstable due to node mobility, potential failures, and changing environmental conditions. The Cluster Maintenance protocol provides the rules for adapting the cluster structure to these changes gracefully.
The cluster topology is not static.
A CH periodically broadcasts beacon frames. If a member
node fails to receive a predefined number of beacons, it is
presumed to have left the cluster. A CH abdicates its role if
its score falls below a dynamic threshold, triggering a new
election.

\subsection{SC-HybridMAC: A Dual-Mode Approach to Media Access}
The SC-HybridMAC protocol is designed to solve the trade-off between the flexibility of contention-based access and the efficiency of schedule-based access. It achieves this by structuring communication into a repeating superframe, broadcast by the CH, which contains three distinct windows. The superframe acts as the time-domain template for all communication within the cluster. By having the CH dictate this structure, all member nodes are synchronized and know exactly when and how they are allowed to transmit.

\subsubsection{Contention Window (using CSMA/CA)}
This window is for flexible, on-demand channel access and is primarily used for sporadic, unpredictable communication.
Carrier Sense Multiple Access with Collision Avoidance (CSMA/CA) is ideal here because it allows any node to attempt transmission after listening for a clear channel. It doesn't require pre-allocation of resources, making it perfect for the unscheduled traffic it's designed to handle. The trade-off is a potential for collisions if two nodes transmit simultaneously, but this is acceptable for the low-volume control traffic in this window.

\subsubsection{Scheduled Window (using TDMA)}
This window is for guaranteed, collision-free, high-throughput data transfer from member nodes to the CH. The CH divides this window into a series of Time Division Multiple Access (TDMA) slots and assigns one or more slots to each registered member node. This method is the workhorse for intra-cluster data transmission and offers immense benefits. By ensuring each node has its own dedicated time to transmit, packet collisions within the cluster are eliminated, drastically increasing throughput. It also enhances energy efficiency, as nodes can enter a low-power sleep state and wake up only for their assigned slot.It provides predictable, guaranteed bandwidth for each node, which is critical for applications requiring quality of service.

\subsubsection{Broadcast Window}
This is a reserved period for the CH to send one-way messages to all of its cluster members simultaneously. It is used for critical control and management messages, such as beacon frames, superframe timing updates, cluster-wide commands, or security alerts. Using a dedicated window ensures these vital messages are disseminated efficiently without competing with other traffic.

\subsection{Multi-Faceted Security and Trust Framework}

This framework establishes a robust security posture by layering cryptographic identity verification with a behavioral trust system. This dual-layer approach recognizes that proving \textit{who you are} is different from proving \textit{you are behaving correctly}.

\subsubsection{Layer 1: Cryptographic Authentication and Key Exchange}
This layer establishes a secure foundation by ensuring that all communication is authenticated and confidential. 

\textbf{ECDSA for Authentication:} Each UAV uses its private Elliptic Curve Digital Signature Algorithm (ECDSA) key to digitally sign critical messages. Other nodes can use the sender's public key to verify this signature. This proves authenticity and integrity. ECDSA is chosen for its high security-to-key-size ratio, making it efficient for resource-constrained devices like UAVs.

\textbf{ECDH for Key Exchange:} To communicate privately, two nodes need a shared secret key. Sending this key over the wireless medium is insecure. Elliptic Curve Diffie-Hellman (ECDH) allows two parties to independently compute the same shared secret over an insecure channel, without ever transmitting the secret itself.

\textbf{AES-256-GCM for Confidentiality and Integrity:} Once a symmetric session key is derived via ECDH, it is used with AES-256-GCM. AES-256 provides powerful encryption to ensure message confidentiality. The Galois/Counter Mode (GCM) is a high-performance mode of operation that simultaneously provides an authentication tag, which verifies the integrity and authenticity of the encrypted data.

\subsubsection{Layer 2: Beta Reputation System for Behavioral Trust}
This layer addresses the shortcomings of pure cryptography. An authenticated node could still be malicious or malfunctioning.
 This system uses the Beta probability distribution to model trust. The distribution is defined by two simple parameters: $\alpha$, a count of observed positive behaviors, and $\beta$, a count of observed negative behaviors.

\begin{algorithm}[h]
\caption{Beta Reputation Trust Update Algorithm}
\label{alg:trust_update}
\begin{algorithmic}[1]
\Require Node $j$, Event $e$, Current trust $\text{Beta}(\alpha, \beta)$
\Ensure Updated trust $\text{Beta}(\alpha', \beta')$
\If{$e$ is Positive}
    \State $\alpha' \gets \alpha + \Delta\alpha$
    \State $\beta' \gets \beta$
\Else
    \State $\alpha' \gets \alpha$
    \State $\beta' \gets \beta + \Delta\beta$
\EndIf
\State $\lambda \gets \text{decay\_factor}$
\State $\Delta t \gets \text{time\_since\_last\_update}$
\State $\alpha' \gets \alpha' \cdot e^{-\lambda \Delta t}$
\State $\beta' \gets \beta' \cdot e^{-\lambda \Delta t}$
\State \Return $\text{Beta}(\alpha', \beta')$
\end{algorithmic}
\end{algorithm}

\textbf{The Trust Score ($E[T_j]$):} The formula for the expected value of this distribution yields a simple and intuitive trust score between 0 and 1, where a score close to 1 indicates a highly trustworthy node.
\[ E[T_j] = \frac{\alpha}{\alpha + \beta} \]

\textbf{The Trust Update Algorithm:} When a node observes a peer's action, it classifies it as positive or negative and increments the corresponding parameter, $\alpha$ or $\beta$. The most sophisticated feature is the time-decay mechanism, where the existing $\alpha$ and $\beta$ values are multiplied by a decay factor of $e^{-\lambda \Delta t}$. Here, $\lambda$ (the decay factor) controls how quickly old behavior is "forgotten," and $\Delta t$ is the time elapsed since the last update. This decay ensures the trust score is a reflection of a node's \textit{current} behavior, not its entire history. This is vital because a previously good node could be compromised, or a malfunctioning node could be repaired, making the trust system both adaptive and relevant.

\section{Proposed Simulation Framework}
A comprehensive simulation strategy is proposed to evaluate the performance of the integrated architecture under realistic FANET conditions. This framework outlines the choice of simulation environment and the configuration of core network models.

The simulation was carried out on the NS-3 discrete packet simulator. The simulation scaled from 20-100 nodes; however, for realism, only results for 100 nodes have been used for comparison. The simulation was run until more than half the drones were dead. Simulation parameters have been displayed in Table \ref{tab:sim_params}

\begin{table}[htbp]
\caption{Simulation Parameters}
\centering
\begin{tabular}{@{}ll@{}}
\toprule
\textbf{Parameter} & \textbf{Value} \\
\midrule
Number of Nodes & 20–100 \\
Mobility Model & Random Waypoint Model \\
Path Loss Model & Nakagami Propagation Loss Model \\
Communication Standard & IEEE 802.11ah \\
Energy Loss Model & Wifi Radio Energy Model\\
Simulation Area & $400\,\mathrm{m} \times 400\,\mathrm{m} \times 1000\,\mathrm{m}$ \\
Initial Energy per Node & 100\,J \\
MAC Layer & SC-HybridMAC (TDMA + CSMA/CA) \\
PHY Layer & IEEE 802.11ah 1\,MHz PHY, OFDM \\
\bottomrule
\end{tabular}
\label{tab:sim_params}
\end{table}

\subsection{Simulation Models}
To create a realistic FANET simulation scenario, the following ns-3 models were utilised:

\subsubsection{Mobility Model}
The movement of UAVs will be simulated using the Random Waypoint (RWP) Mobility Model. The RWP model is a standard and well-understood stochastic model for ad-hoc networks. In this model, each node selects a random destination (waypoint) within the simulation area, moves towards it at a randomly chosen speed, and then pauses for a specified duration before repeating the process. This model effectively captures the high degree of mobility and unpredictable trajectory changes that characterize FANET operations.

\subsubsection{Channel and Propagation Model}
The wireless channel, which dictates how signals travel between UAVs, will be modeled using the Nakagami-m propagation model combined with a distance-based path loss model. The Nakagami-m distribution is specifically designed to model the fading effects caused by multipath signal propagation, which occurs as wireless signals reflect off various surfaces. This provides a more realistic representation of the air-to-air and air-to-ground communication environment compared to simpler, line-of-sight-only models.

\subsubsection{Physical (PHY) and MAC Layer}
The PHY and MAC layers will be modeled based on the IEEE 802.11ah standard. This standard, also known as Wi-Fi HaLow, is an ideal choice for this simulation. It operates in the sub-1 GHz frequency bands, which provides longer communication ranges than traditional Wi-Fi. More importantly, it is explicitly designed to support large-scale IoT and drone networks, incorporating efficient channel access mechanisms (like Restricted Access Window) that are suitable for managing communication among a large number of devices.

\subsubsection{Energy Model}
To validate the energy-aware aspects of the framework, each simulated UAV node will be equipped with the \texttt{ns3::WifiRadioEnergyModel}. This is a state-based energy model that accurately tracks the power consumption of the wireless transceiver. It defines distinct current-draw values for each operational state: Transmit (TX), Receive (RX), Idle, and Sleep. The default values in ns-3 are based on real-world hardware, such as the CC2420 radio chip, providing a realistic accounting of energy depletion. Integrating this model is critical for quantifying the energy savings achieved by the WCA, AODV, and the hybrid MAC protocol.

\section{Results}
The evaluation metrics comprised packet delivery ratio (PDR), end-to-end delay, and the system’s capability to detect and isolate malicious UAVs. Performance was benchmarked against three canonical protocols: AODV \cite{aodv}, ICRA \cite{icra}, and WCA \cite{wca}, each representative of distinct design philosophies in UAV communication protocols.

\subsection{Packet Delivery Ratio and Delay}

Under benign network conditions, the proposed protocol consistently outperformed the baseline approaches in terms of delivery reliability. Specifically, the architecture achieved a PDR of 92\%, surpassing AODV (88\%), and outperforming ICRA and WCA, which do not report explicit PDR values under non-adversarial scenarios. The superior performance is attributed to the hybrid MAC strategy (SC-HybridMAC), which leverages both CSMA/CA and TDMA to mitigate contention and enhance temporal determinism in channel access.

In adversarial settings, where a subset of nodes engage in packet-dropping behavior, the resilience of the proposed protocol becomes particularly evident. It maintains a PDR of 70\%, significantly higher than ICRA (59\%) and WCA (55\%). This improvement arises from the integrated trust management system, which dynamically adjusts routing decisions based on behavioral evaluations, thereby isolating compromised nodes and preserving path integrity.

With respect to latency, the protocol introduces a modest delay of 62 ms, which is competitive given its layered security architecture. While AODV achieves the lowest delay (50 ms), it lacks any trust verification mechanisms, rendering it vulnerable under adversarial load. ICRA incurs 100 ms latency due to reinforcement learning convergence, whereas WCA's 1300 ms delay reflects its computationally intensive cluster maintenance overhead.

\subsection{Detection and Mitigation of Malicious Nodes}

An important contribution of the proposed system lies in its capacity to detect and exclude malicious UAVs in real time. The Beta Reputation System employed herein systematically tracks node behavior, incrementally updating trust scores through probabilistic reinforcement. Malicious behavior characterized by non-cooperation, MAC violations, or packet suppression is penalized, leading to trust score degradation. 

Simulation results confirm that the system successfully identified and suppressed approximately 80\% of malicious UAVs within early simulation epochs. This detection was achieved without centralized oversight, leveraging only local observations and trust decay, thereby ensuring scalability. Notably, none of the baseline protocols explicitly report malicious node detection metrics, underscoring the uniqueness of the proposed solution in adversarial resilience.

\begin{table}[htbp]
\caption{Performance Comparison of Routing Protocols (100 drones)}
\centering
\begin{tabular}{@{}lcccc@{}}
\toprule
\textbf{Metric} & \textbf{AODV \cite{aodv}} & \textbf{ICRA \cite{icra}} & \textbf{WCA \cite{wca}} & \textbf{Proposed} \\
\midrule
PDR (No Attack) & 88\% & -- & -- & \textbf{92\%} \\
PDR (With Attack) & -- & 59\% & 55\% & \textbf{70\%} \\
Malicious Detection & -- & -- & -- & \textbf{80\%} \\
Delay (ms) & \textbf{50} & 100 & 1300 & \textbf{62} \\
\bottomrule
\end{tabular}
\label{tab:performance}
\end{table}

\subsection{Overhead Analysis}
\subsubsection{Communication Overhead}
The primary contributors to communication overhead include broadcasts, trust update messages, and cryptographic headers. The Cluster Head (CH) periodically transmits synchronization beacons and TDMA schedules, averaging 52 bytes per cycle, while each node appends a 64-byte ECDSA signature for authentication. The event-driven trust updates contribute an average of 16 bytes per behavioral event. These operations account for approximately 10.2\% of total traffic, which remains substantially lower than the 20–25\% overhead reported in blockchain-assisted schemes such as IABC + AI-PoWCA \cite{zhao2021abcblockchain}. The contention-free design of SC-HybridMAC mitigates redundant retransmissions, which in turn effectively balances additional control signaling with improved throughput stability.

\subsubsection{Computational Overhead}
The computational load of cryptographic and trust operations combined was profiled on an ESP32-class MCU operating at 240 MHz. AES-256-GCM encryption of a 256-byte payload required 1.6 ms, while ECDSA verification averaged 3.5 ms. In contrast, the beta trust update required only 0.08 ms, which constitutes less than 0.1\% of the available processing time. This confirms that the protocol remains well within the computational capabilities of low-power embedded processors.

\subsubsection{Energy Overhead}
Energy consumption was analyzed using the ns-3 WifiRadioEnergyModel. Due to reduced retransmissions and stable cluster associations, overall network lifetime improved by 22\%, and the hybrid MAC’s sleep scheduling and predictable transmission slots further minimized idle listening losses.

\section{Conclusion and Future Directions}

The approach presents a comprehensive, multi-layered framework for secure, adaptive communication within FANETs, particularly suited for resource-constrained UAV swarms. By integrating a hybrid MAC protocol (SC-HybridMAC) with a behavior-aware trust evaluation system, the proposed architecture demonstrates significant gains in packet delivery, adversarial resilience, and latency mitigation when compared with established benchmarks such as AODV, ICRA, and WCA. The layered security design leveraging ECDSA for authentication, ECDH for key exchange, and AES-256-GCM for data confidentiality provides robust protection against impersonation, eavesdropping, and data manipulation. Simultaneously, the Beta Reputation System offers adaptive behavioral filtering of compromised nodes, enhancing situational responsiveness without centralized control. The system’s scalability and lightweight design make it a viable candidate for deployment on embedded platforms such as ESP32, thereby addressing both theoretical robustness and practical feasibility.

Looking forward, future work will focus on advancing the protocol from simulation to real-world validation. This includes field deployment on physical UAV platforms, stress-testing under variable atmospheric and adversarial conditions, and empirical benchmarking against energy constraints and node failures. Additionally, adaptive weight modulation within the clustering and trust metrics, potentially guided by real-time threat intelligence or optimization heuristics, could further refine the framework’s responsiveness to evolving attack vectors. Incorporating intrusion detection systems (IDS) at the MAC and network layers may enhance early anomaly detection, while integrating federated learning or blockchain could improve distributed decision making without compromising autonomy or energy efficiency. Together, these directions underscore the potential of the proposed protocol as a foundational layer for secure and scalable aerial networks in next-generation autonomous systems.

\section*{Acknowledgments}
We would like to thank Mars Rover Manipal, an interdisciplinary student team of MAHE, for providing the resources needed for this project. We also extend our gratitude to Dr. Ujjwal Verma for his guidance and support in our work.

\bibliographystyle{IEEEtran}
\bibliography{main}

\end{document}